\documentclass[10pt,twocolumn]{article}
\usepackage[top=1.9cm, bottom=1.9cm, left=1.8cm, right=1.8cm]{geometry}
\usepackage{times}  %
\usepackage[sort&compress,numbers]{natbib}  %
\usepackage[hyphens]{url}
\usepackage{graphicx}  %
\usepackage[keeplastbox]{flushend}
\usepackage[hyphens]{url}  %
\usepackage{graphicx} %
\usepackage{tikzsymbols}

\newcommand{\descr}[1]{\smallskip\noindent\textbf{#1.}}

\usepackage{sectsty}
\sectionfont{\bfseries\Large\raggedright}

\usepackage{url}                                %
\usepackage{array,multirow,graphicx,adjustbox}  %
\usepackage{booktabs}                           %
\usepackage[utf8]{inputenc}
\usepackage[ruled,algosection,noend,linesnumbered]{algorithm2e}
\usepackage{float}
\usepackage{paralist}
\usepackage{amsmath}
\usepackage{amsfonts}
\usepackage{xcolor}
\usepackage{csquotes} 
\usepackage{tabularx}
\usepackage{makecell}
\usepackage{pbox}
\usepackage[hang,flushmargin]{footmisc}
\usepackage{flushend}
\usepackage{xspace}
\usepackage{multicol, lipsum}
\usepackage[T1]{fontenc}

\usepackage{xcolor}
\usepackage{booktabs}
\usepackage{graphicx}
\usepackage{paralist}
\usepackage[bf]{caption}
\usepackage{subcaption}

\let\oldbibliography\thebibliography
\renewcommand{\thebibliography}[1]{%
  \oldbibliography{#1}%
  \setlength{\itemsep}{2pt}%
}

\usepackage[hang,flushmargin]{footmisc}

\usepackage[compact]{titlesec}
\titlespacing*{\section}{0pt}{*4}{4pt}
\titlespacing*{\subsection}{0pt}{*2.5}{2.5pt}
\usepackage{xspace}

\makeatletter
\def\url@leostyle{%
  \@ifundefined{selectfont}{\def\UrlFont{}}%
  {\def\UrlFont{}}%
}
\makeatother
\usepackage[hyphenbreaks]{breakurl}

\usepackage[bookmarks=true, bookmarksnumbered=true, colorlinks=true, linkcolor=linkcol, citecolor=citecol, urlcolor=urlcol, hypertexnames=true]{hyperref}

\definecolor{darkgreen}{RGB}{0, 100, 0}
\definecolor{linkcol}{rgb}{0.3,0,0}
\definecolor{citecol}{rgb}{0.3,0,0}
\definecolor{urlcol}{rgb}{0.3,0,0}

\makeatletter
\def\url@leostyle{%
  \@ifundefined{selectfont}{\def\UrlFont{\small}}%
  {\def\UrlFont{}}%
}
\makeatother
\newcommand{\jbnote}[1]{}
\newcommand{\synote}[1]{}
\newcommand{\edc}[1]{}

\begin{document}

\title{\bf Understanding the Use of e-Prints on Reddit and 4chan's Politically Incorrect Board\thanks{Published in the Proceedings of the 15th ACM Web Science Conference 2023 (ACM WebSci 2023). Please cite the WebSci version.}}

\author{
	Satrio Baskoro Yudhoatmojo$^1$,
	Emiliano De Cristofaro$^2$,
	Jeremy Blackburn$^1$ \\[0.5ex]
  $^1$Binghamton University, $^2$University College London
}

\date{}

\maketitle

\begin{abstract}
The dissemination and reach of scientific knowledge have increased at a blistering pace.
In this context, e-Print servers have played a central role by providing scientists with a rapid and open mechanism for disseminating research without waiting for the (lengthy) peer review process.
While helping the scientific community in several ways, e-Print servers also provide scientific communicators and the general public with access to a wealth of knowledge without paying hefty subscription fees.
This motivates us to study how e-Prints are positioned within Web community discussions.

In this paper, we analyze data from two Web communities: 14 years of Reddit data and over 4 from 4chan's Politically Incorrect board.
Our findings highlight the presence of e-Prints in both science-enthusiast and general-audience communities.
Real-world events and distinct factors influence the e-Prints people's discussions; e.g., there was a surge of COVID-19-related research publications during the early months of the outbreak and increased references to e-Prints in online discussions.
Text in e-Prints and in online discussions referencing them has a low similarity, %
suggesting that the latter are not exclusively talking about the findings in the former.
Further, our analysis of a sample of threads highlights: 1)~misinterpretation and generalization of research findings, 2)~early research findings being amplified as a source for future predictions, and 3)~questioning findings from a pseudoscientific e-Print.
Overall, our work emphasizes the need to quickly and effectively validate non-peer-reviewed e-Prints that get substantial press/social media coverage to help mitigate wrongful interpretations of scientific outputs. 
\end{abstract}

\section{Introduction}
One of the main products of scientific research is the publication of articles and papers.
In fact, this is explicitly built into most academic career paths~\cite{Wang_2020}, as per the infamous ``publish or perish'' promotion paradigm.
This translates into intense competition both in terms of cutting-edge research, but also in the quick and broad dissemination of findings.
In this context, e-Print servers, e.g., arXiv, bioRxiv, and medRxiv, provide scientists with a rapid and open mechanism for disseminating their work. 

Besides enabling researchers to spread the word quickly and elicit feedback about their research (or even simply ``timestamp'' it), e-Prints are also increasingly pivotal in science communication and outreach.
Science journalists, policymakers, and enthusiasts can easily monitor and search for articles on a handful of e-Print servers without paywalls or subscription fees.
Laypeople can do the same, possibly following links from press coverage.

Articles posted on e-Print servers are often not peer-reviewed, and there is essentially no quality control on what is published.
As a result, users may accidentally or intentionally be exposed to questionable papers.
This issue is further compounded by users sharing possibly flawed, inconclusive, or misinterpreted results and treating them as a gold standard.
Consider the explosion of e-Prints in the early months of the COVID-19 pandemic~\cite{Majumder_2020,Vlasschaert_2020,Fleerackers_2021,Fraser_2021}, %
which led to dubious claims and proposed treatments being disseminated on bioRxiv and medRxiv~\cite{Kwon_2020}.

In this paper, we explore both the societal and scientific explorations of openly accessible, possibly non-peer-reviewed, e-Prints.
We set out to better understand how e-Prints are positioned within two distinct Web communities: Reddit and 4chan's Politically Incorrect board (/pol/).
Reddit provides numerous and distinct subcommunities, while 4chan is a hotbed of extremism~\cite{Hine_2017}, which has a substantial impact on Internet culture~\cite{Zannettou_2018}, and significant influence on the spread of disinformation~\cite{Zannettou_2017}.
Our data-driven exploration of these two communities thus provides a balance between more mainstream discussions and a community well known for being outside the mainstream.

\descr{Research Questions} Overall, we focus on three research questions:
\begin{itemize}
    \item[\textbf{RQ1:}] What is the general presence of e-Prints on Reddit and /pol/, and what communities are linking to/discussing them?
    \item[\textbf{RQ2:}] What \emph{kind} of e-Prints are linked to Reddit submissions and /pol/ posts?
    \item[\textbf{RQ3:}]  How are e-Prints discussed in Reddit submissions and /pol/ posts? 
\end{itemize}

To answer these questions, we build on several data-driven methods using a curated dataset of over 14 years of Reddit data, over four years of /pol/ data, and 54,684 e-Prints.
Using an embeddings-based topic model, we explore what kind of e-Prints are posted on Reddit and /pol/.
Next, we train a document embedding model to assess the similarity between the contents of e-Prints and the social media discussion around them.
We then perform a qualitative assessment of Reddit submissions and 4chan threads to gain a deeper knowledge of how the linked e-Prints are being used in discussions.

\descr{Main Findings} Our work highlights several findings:
\begin{enumerate}
  \item We present evidence of e-Print presence on Reddit and /pol/.
        On Reddit, we find arXiv e-Print links are shared the most, with 50,904 e-Prints, followed by viXra (975), bioRxiv (781), ChemRxiv (69), medRxiv (60), and EarthArxiv (24).
        On /pol/, we find medRxiv to be the top shared e-Prints with 719 e-Prints, followed by bioRxiv (676), arXiv (427), viXra (27), PsyArXiv (17) and ChemRxiv (5).
  \item Our topic modeling reveals that COVID-19-related e-Prints from bioRxiv and medRxiv surged in the weeks following the beginning of the pandemic.
  \item We find that social media discussions and linked e-Prints exhibit low similarity -- less than 0.32 on average.
  \item Via manual examination of social media discussions, we find evidence that e-Prints are misinterpreted to fit the position of a community or individual, the conclusions e-Prints are generalized without adequately explaining their findings and context, early research findings being used to speculate on COVID-19 infection rate, and science enthusiasts collectively debunking pseudoscientific e-Prints.
\end{enumerate}

\section{Background and Related Work}
\label{sec:background-and-related-work}

\subsection{e-Prints}
\label{sec:e-Prints}
\descr{{Preprint}} 
A \emph{preprint} is a term used for an unpublished document, e.g., an advanced copy of a scientific article intended for publication in a journal or conference proceeding~\cite{Addis_1970}.
Preprints in science have been used as an informal medium for information exchange to short circuit some barriers in scientific output, e.g., publication delays, increasing visibility compared to the growth of publications, and the competitive nature of academia placing pressure on being ``the first'' to present findings.
The field of particle/high-energy physics was an early pioneer in using preprints, distributing them via physical mail before preprint servers existed~\cite{Addis_1970,Ginsparg_2011,Xie_2021}.

In 1991, Paul Ginsparg at Los Alamos National Laboratory in New Mexico launched an electronic bulletin board to distribute preprints in theoretical High-Energy Physics (hep-th) with a notification system sent using the email address hep-th@xxx.lanl.gov~\cite{Ginsparg_2011}.
Later, the electronic bulletin board evolved into a full-blown website hosted at arXiv.org.
Over time, the preprints on arXiv grew beyond Physics to broader fields like Mathematics, Computer Science, and Statistics (for a complete listing of arXiv categories, see~\cite{arxiv_categories}).

\descr{{e-Print}}
Nowadays, the more appropriate term to describe the broad preprint ecosystem is \emph{e-Print}.
e-Print acknowledges the fact that papers might have been published (or about to be) in peer-reviewed venues, but the authors are also making them open-access.
Other fields have also adopted e-Print culture, with most deriving the name for their servers from the original arXiv server.
Over 40 e-Print servers were created between 2000 and 2010~\cite{Xie_2021}.
Some of these are general-purpose, e.g., PeerJ PrePrints, Preprints.org, and viXra.
There are also servers for specific disciplines (e.g., LawArXiv, SportsArxiv, ChemRxiv, PsyArXiv, SocArXiv, and EarthArXiv) and regional ones (e.g., AfricArXiv, Arabixiv, IndiaRxiv, and Frenxiv).
In recent years, the creation of focused e-Print servers has not diminished.
For instance, bioRxiv, focusing on biology, was launched in 2013~\cite{biorxiv_categories,Hoy_2020}, while medRxiv, focusing on medical sciences, in 2019~\cite{medrxiv_categories,Hoy_2020}.

\descr{{Benefits}}
As a medium for informal information exchange, one advantage of e-Prints is the rapid distribution of the article to potential readers~\cite{Addis_1970}, which allows authors to be the first to publicize/timestamp their findings as well as for potential readers to read and give feedback to the authors as soon as possible~\cite{Hoy_2020}.
As feedback is considered and revisions are made, the e-Print can be updated with a newer version on the server.
Another advantage is that e-Print servers are accessible to the public, unlike typical publishing platforms, which hinder public access via paywalls.
Therefore, e-Prints have an inherently broader audience than articles published in many well-established, peer-reviewed venues that lie behind the paywalls.
Overall, e-Prints allow for early recognition, fast and broad dissemination, open access, and even facilitate collaboration~\cite{Sarabipour_2019,Hoy_2020,Wang_2020}.

\descr{{Drawbacks}}
The other side of the coin is that e-Prints do not undergo the peer review process accompanying publication in traditional venues~\cite{Hoy_2020}.
Thus, there is no real control (except for post-review) over the quality of e-Prints, yet they are still regularly cited by other scientific work.
Although reducing academic gatekeeping is a laudable goal, at minimum, the peer review process provides some level of assurance that a paper has at least been looked at by scientists who are not the authors.
e-Prints also contribute to information overload.
Because their number is essentially unbounded, scientists must consider a substantially larger body of literature when exploring their problem domain.
Again, while this is not necessarily bad, it has consequences; the larger the haystack, the more difficult it is to find the needle hidden within.

\subsection{Related Work}
\label{sec:related-work}

\descr{{e-Prints}}
Xie et al.~\cite{Xie_2021} study the exponential growth of e-Prints over 30 years, finding that arXiv is by far the largest.
Two-thirds of bioRxiv e-Prints posted before 2017 ended up in peer-reviewed journals~\cite{Abdill_2019}, while 30\% of bioRxiv e-Prints remain unpublished in peer-reviewed venues, and half of the published ones ended up in Elsevier, Nature, PLOS, and Oxford University journals~\cite{Anderson_2020}.
A study on Computer Science e-Prints posted on arXiv between 2008 and 2018 shows that peer-reviewed e-Prints differ in several ways from their published version, with, e.g., changes to titles, authorship, abstract and introduction, the addition of more authoritative references, and the availability of source code~\cite{Lin_2020}.
Klein et al.~\cite{Klein2016} measure the similarity between pre-Print papers and their final published counterparts, revealing that most final published papers are similar to their pre-Print versions.
A study on the main driving factors in accelerating scientific knowledge dissemination shows the early-view and open-access effects of the e-Prints contribute to measurable citations and readership, as well as visibility~\cite{Wang_2020}.

The recent COVID-19 outbreak has driven a surge of scientific papers in which e-Print servers are influential in allowing them to timestamp the findings and disseminate them to the public quickly.
Moreover, scientific communications on social media and Web communities that mention e-Prints have increased during the pandemic.
Vlasschaert et al.~\cite{Vlasschaert_2020} argue that the scientific community should take advantage of e-Prints and embrace the culture of open and critical discussions instead of depending solely on the peer review process as a quality control mechanism.
A content analysis study of COVID-19-related e-Prints usage by media outlets reveals that traditional news outlets and news aggregators heavily covered them during the early months of the outbreak when our understanding of the pandemic was in its infancy~\cite{Fleerackers_2021}.
COVID-19 e-Prints also receive increased scientific and public engagement, with shorter review times and widespread use by journalists and policymakers~\cite{Fraser_2021}.

\descr{{Scientific communication}}
Several studies have investigated scientific communication on Reddit.
For instance, the r/science subreddit provides substantial information exchange, and the comments produced a unique science communication that guides engagement with scientific research~\cite{Jones_2019}.
Also, the frequent contributors of r/science use specialized language to discuss the research findings, as opposed to transient contributors and contributors that eventually left r/science, implying that technical language serves as a gatekeeper to prevent contributors whose language is not aligned with frequent contributors~\cite{August_2020}.
Kousha and Thelwall~\cite{Kousha_2020} assess the coverage of scholarly databases and impact indicators between March 11 to April 18, 2020, and find that increasing research publications are more accessible through Dimensions database compared to Scopus, Web of Science, and PubMed.
In fact, a few COVID-19 publications listed in~\cite{Kousha_2020} have gained substantial attention in the news and social media.

\descr{{Social media studies}}
Hine et al.~\cite{Hine_2017} show that many links posted on /pol/ are ``right-wing'' news sources and shed light on ``raiding'' behavior where /pol/ users would go to YouTube to post hate in video comments.
Zanettou et al.~\cite{Zannettou_2017} show that alt-right communities on /pol/ and Reddit significantly influenced Twitter in propagating ``alternative'' news to mainstream social networks.
Another study by Zannettou et al.~\cite{Zannettou_2018} focuses on weaponized memes by analyzing the propagation, evolution, and influence of Internet memes on Twitter, Reddit, /pol/, and Gab.
Grover et al.~\cite{Grover_2019} uncover alerting behavior of individual extremists in an online environment through behavioral text pattern analysis of a radical right-wing community on Reddit (i.e., r/altright).

By studying the characteristics of user actions in the threads of r/politics and r/worldnews subreddits, Guimar{\~a}es et al.~\cite{Guimaraes_2019} classify different patterns of controversies into disputes, disruptions, and discrepancies.
LaViolette et al.~\cite{LaViolette_2019} look at r/MensRight and r/MensLib subreddits and shed light on their ideological differences using text classifiers, keyword frequencies, and qualitative approaches.
Aldous et al.~\cite{Aldous_2019} develop a prediction model to predict whether an article will be shared on another social media platform by evaluating how topics affect audiences across five social media platforms (Facebook, Instagram, Twitter, YouTube, and Reddit) at four levels of engagement, achieving 80\% precision.

Motivated by the emergence of ``direct-to-consumer'' genetic testing, Mittos et al.~\cite{Mittos_2020,mittosAnalyzingGeneticTesting2020} analyze the highly toxic language used in the genetic testing discussion on Reddit and /pol/.
Rajadesingan et al.~\cite{Rajadesingan_2020} show that pre-entry learning of the norm stability contributed the most to maintaining stable ``toxic'' norms on political subreddits.
That is, newcomers' comments tend to be different from the behavior of the same people on other subreddits.
Therefore, behavior adjustments are community-specific and not broadly transformative.
 
By building word embeddings of the discussions on Reddit, Ferrer et al.~\cite{Ferrer_2021} uncover gender bias, religious bias, and ethnic bias.
Guimar{\~a}es et al.~\cite{Guimaraes_2021} design a feature space and implement a classifier to predict a controversial post-event given a prefix of a path to a Reddit discussion thread (i.e., US politics, World Politics, Relationships, and Soccer).
Rajadesingan et al.~\cite{Rajadesingan_2021} discover abundant political talk in non-political subreddits with less toxic comments.
Veselovsky et al.~\cite{Veselovsky_2021} use neural embedding to measure the social and cultural context of large-scale online music sharing on Reddit and find a large amount of online music sharing was driven by extra-musical factors, e.g., if the artist is associated with meme culture.
Orii et al.~\cite{Orii_2021} study the sentiment of truckers on the r/Truckers subreddit towards the impact of autonomous trucks on the trucking industry using qualitative method and find only 0.98\% of the comments had positive views on automation.

\begin{table*}[t!]
  \centering
  \small
    \setlength{\tabcolsep}{4pt}
  \begin{tabular}{l|rrr|rrrr}
      \toprule
\multicolumn{1}{c}{} &   \multicolumn{3}{c}{\textbf{\em /pol/}}  & \multicolumn{4}{|c}{\textbf{\em Reddit}}  \\
                        \midrule
      \textbf{Server}     & \textbf{\#Posts} & \textbf{\#Links} & \textbf{\#e-Prints} & \textbf{\#Submissions} & \textbf{\#Comments} & \textbf{\#Links} & \textbf{\#e-Prints} \\
      \midrule
      arXiv      & 2,422 & 525 & 427 & 42,749 & 66,610 & 51,108 & 50,904 \\
      biorRxiv   & 3,793 & 898 & 676 & 785    & 490    & 874    & 781 \\
      medRxiv    & 4,161 & 885 & 719 & 43     & 170    & 70     & 60 \\
      ChemRxiv   & 452   & 11  & 5   & 112    & 383    & 88     & 69 \\
      viXra      & 189   & 44  & 27  & 1,161  & 964    & 1,084  & 975 \\
      PsyArXiv   & 40    & 19  & 17  & 0      & 0      & 0      & 0 \\
      EarthArxiv & 0     & 0   & 0   & 23     & 32     & 25     & 24 \\
      SocArxiv   & 0     & 0   & 0   & 0      & 0      & 0      & 0 \\
      \midrule
      \textbf{Total} & \textbf{11,057} & \textbf{2,382} & \textbf{1,871} & \textbf{44,873} & \textbf{68,649} & \textbf{53,249} & \textbf{52,813} \\
      \bottomrule
  \end{tabular}
  \caption{e-Print server statistics on /pol/ and Reddit.}
  \label{tab:dataset-statistic}
  
\end{table*}

\section{Dataset}
In this section, we report on our data collection methodology and provide an overview of the dataset we use in our study.

\descr{{Reddit}}
Reddit is a social news aggregation and discussion website where posts created by a user can be up-voted or down-voted by others~\cite{Reddit_2022}.
Comments on a post can be replied to and be up- or down-voted.
Reddit has sub-communities called \emph{subreddits}, and each subreddit is associated with a particular area of interest (e.g., politics, movies, science).

We acquire Reddit data from Pushshift~\cite{baumgartner_2020}.
More precisely, we collect 1)~Reddit submission data spanning from June 1, 2005, to March 16, 2021 and 2)~Reddit comment data spanning from December 1, 2005, to March 16, 2021.
We filter and keep the submissions and comments containing links to e-Print servers using simple regular expression match and continue removing links to e-Print servers' homepages, search interface, author page, etc.

\descr{{4chan's Politically Incorrect Board}}
4chan is an image-sharing bulletin board where anyone can share images and post comments anonymously without creating an account~\cite{4chan_2022}.
It consists of many boards with particular interest focus, and each board has many threads discussing similar interest themes as the board.
Moreover, the threads are known for their ephemerality (i.e., threads are continuously deleted).
One particular board that we focus on in this study is the ``Politically Incorrect'' board or /pol/; this is unique as it performs no moderation at all--``everything goes.''
Many discussions in /pol/ are close to far-right and alt-right movements and exhibit xenophobia, social conservatism, racism, and hate~\cite{Hine_2017}.

We acquire /pol/ data from Papasavva et al.~\cite{Papasavva_2020}, with threads from June 30, 2016, to March 16, 2021.
A similar filtering task to Reddit data is applied to /pol/ data to filter and keep threads containing links to e-Print servers.

\descr{{e-Print}}
The e-Print articles' links are extracted from Reddit and /pol/ data.
Mainly, we are interested in links to eight e-Print servers: 1)~arXiv, 2)~bioRxiv, 3)~medRxiv, 4)~ChemRxiv, 5)~PsyArXiv, 6)~viXra, 7)~EarthArxiv, and 8)~SocArxiv.
These eight servers do not have a peer review system.
However, they do have a moderation system to screen the submissions.

\begin{figure*}[t!]
  \centering
  \begin{subfigure}{0.495\linewidth}
      \centering
      \includegraphics[width=\linewidth]{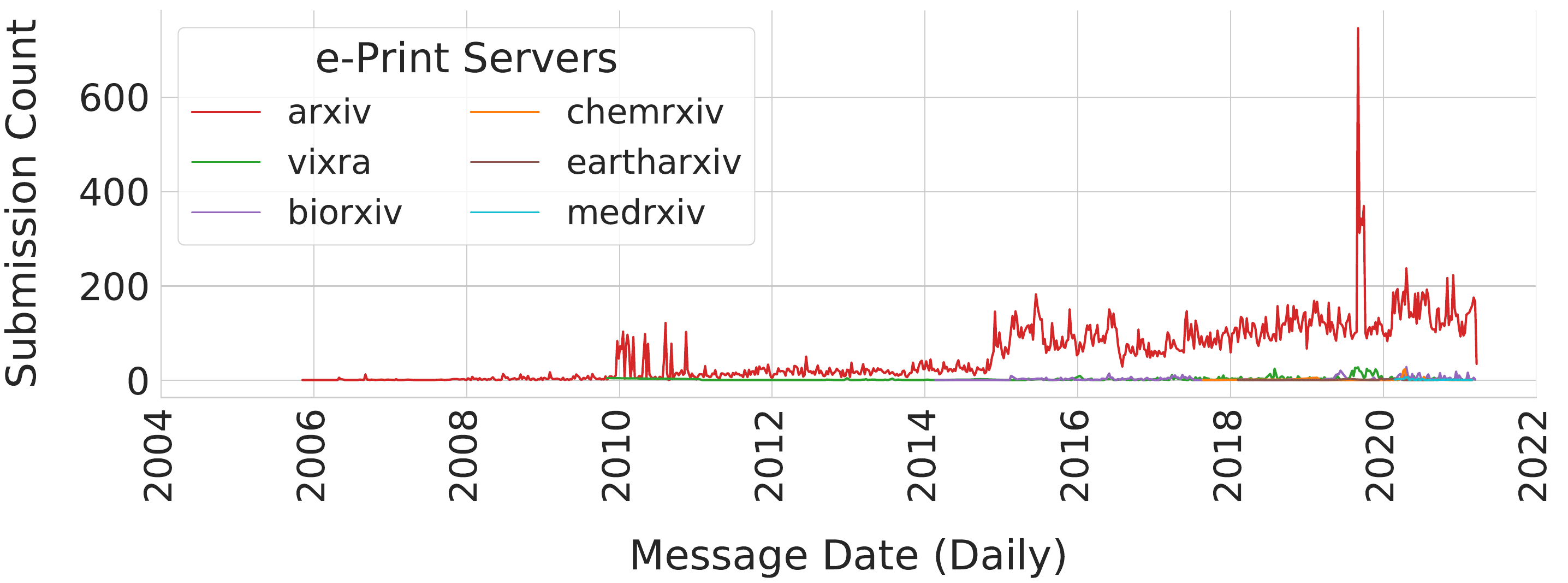}
      \caption{\#Reddit {\em submissions} with e-Print links} %
      \label{fig:reddit-submissions-count}
  \end{subfigure}
  \hfill
  \begin{subfigure}{0.495\linewidth}
      \centering
      \includegraphics[width=\linewidth]{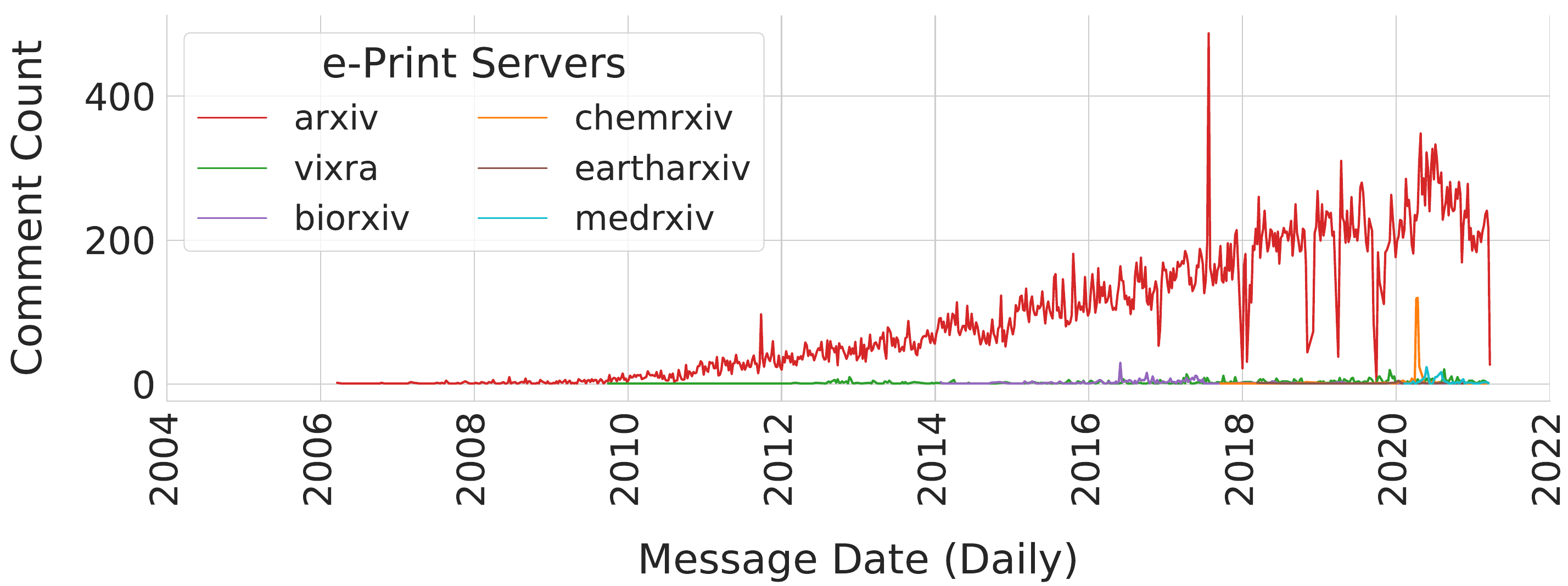}
      \caption{\#Reddit {\em comments} with e-Print links} %
      \label{fig:reddit-comments-count}
  \end{subfigure}
  \medskip
  \begin{subfigure}{0.495\linewidth}
      \centering
      \includegraphics[width=\linewidth]{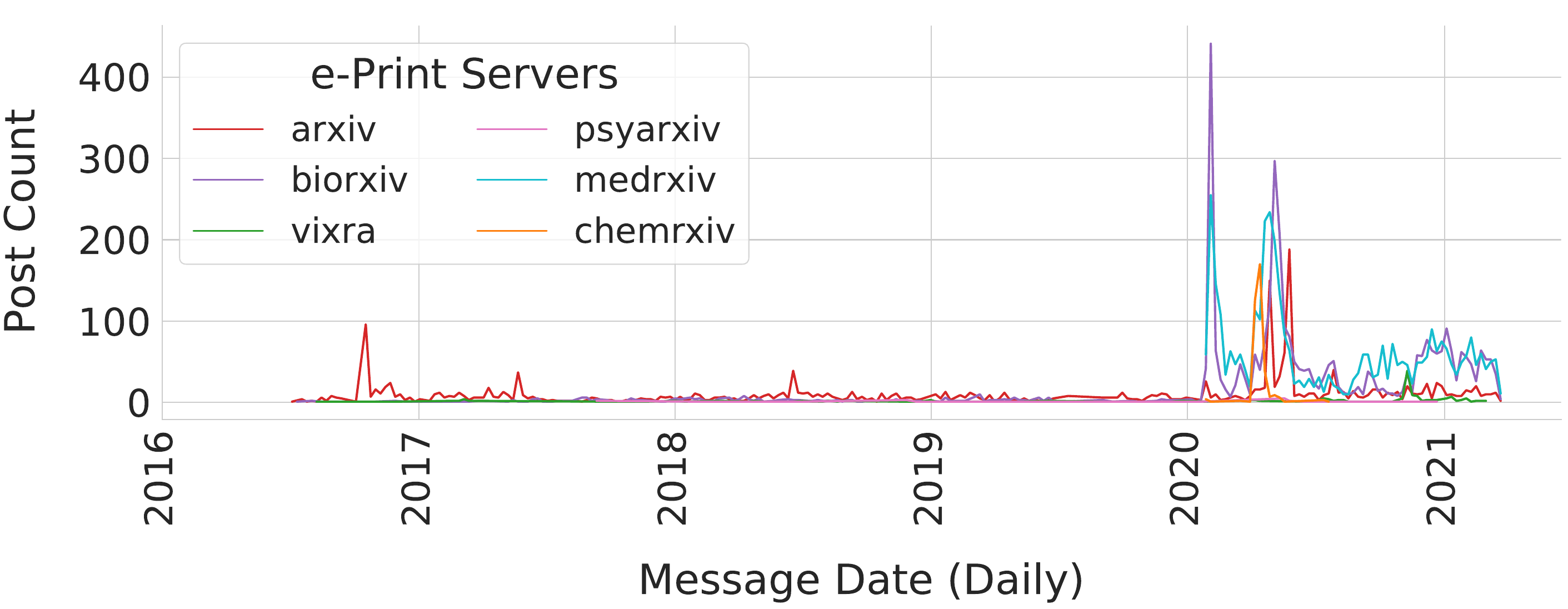}
      \caption{\#/pol/ posts with e-Prints links}
      \label{fig:pol-posts-count}
  \end{subfigure}
  \caption{Reddit submissions/comments and /pol/ posts containing links to e-Prints.}
  \label{fig:posts-count}
\end{figure*}

\descr{Ethical Considerations}
We understand that data from Web communities like Reddit and 4chan may contain personal information; therefore, we use standard best practices to ensure we follow basic ethical principles~\cite{bailey2012menlo,rivers2014ethical}.
That said, we \emph{did} perform analysis that was not purely aggregate in fashion via our case studies.
In our case studies, we present some examples taken from Reddit and /pol/ posts.
However, we paraphrased the posts to mitigate disclosing the original post and the users who posted the content.
Moreover, we do not include any usernames in our paraphrased quotes.
Therefore, our work does not pose any substantial ethical concerns.

\section{Results}

\subsection{RQ1: What is the general presence of e-Prints on Reddit and /pol/?}

\descr{Overview} To uncover the general presence of e-Prints on Reddit and /pol/, we measure, temporally, the distribution of e-Print links and engagement of e-Prints on Reddit submissions and comments and /pol/ threads.
We count the daily number of Reddit submissions and comments and /pol/ threads containing each of the eight e-Print servers to present the temporal presence of e-Print links on Reddit and /pol/.
Then, we identify subreddits having the highest number of links to e-Prints.

\descr{General statistics} In Table~\ref{tab:dataset-statistic}, we provide some general statistics about the e-Prints we find on /pol/ and Reddit, more precisely
the number of posts (on the former) and submissions/comments (on the latter) that include a link to an e-Print server, as well as the number of unique links to an e-Print server (\#links) and that of links to an e-Print article (\#e-Print).
Then, in Figure~\ref{fig:posts-count}, we plot the daily number of submissions and comments on Reddit and posts on /pol/ with e-Print links.

Overall, we observe a non-negligible presence of e-Prints in both communities, although no links to PsyArXiv on Reddit, EarthArxiv on /pol/, or SocArxiv on either.

\begin{table}[t!]
  \centering
  \small
  \begin{tabular}{lrrr}
      \toprule
      \textbf{Data} & \textbf{Max}  & \textbf{Min}  & \textbf{Median} \\
      \midrule
      Submissions   & 3,762         & 1             & 1 \\  
      Comments      & 7,643         & 1             & 1 \\
      \bottomrule
    \end{tabular}
  \caption{The statistics of the total number of submissions/comments containing e-Print links in a subreddit.}
  \label{tab:subreddit-stats}    
\end{table}

\begin{figure}[t!]
  \centering
  \includegraphics[width=.83\linewidth]{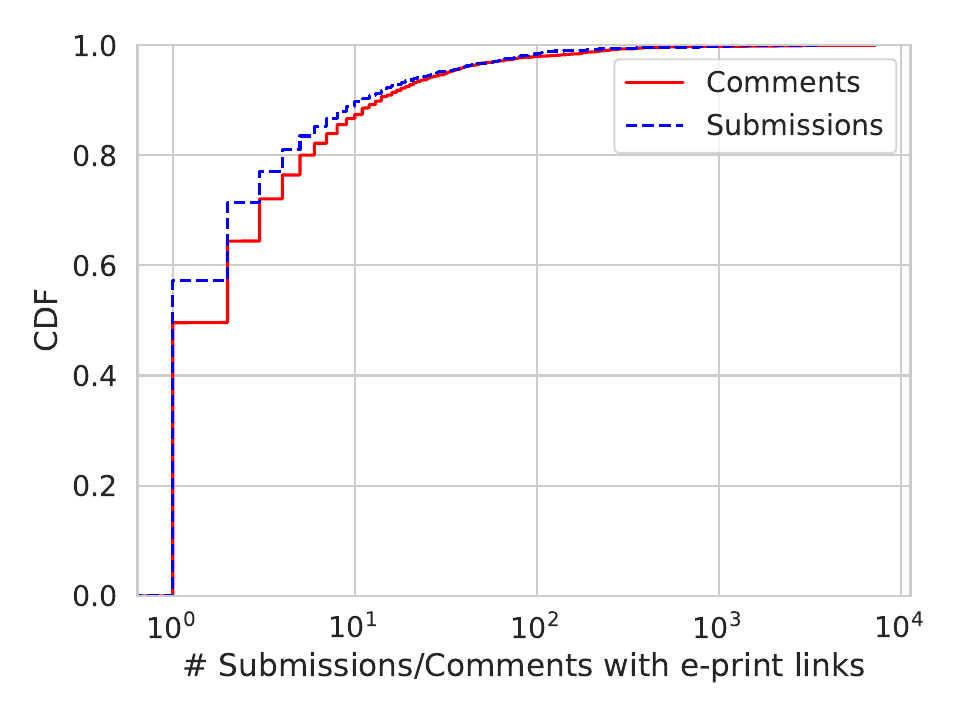}
  \caption{CDF of the total number of submissions/comments containing e-Print links in a subreddit.}
  \label{fig:subreddit-cdf-count}
\end{figure}

\descr{Reddit}
To understand whether links appear throughout or are mostly concentrated on a handful of discussions, in Figure~\ref{fig:subreddit-cdf-count}, we plot the CDF of the number of e-Print links in Reddit submissions/comments.
This shows that about half of subreddits only have a single submission or comment with an e-Print link.

Next, in Table~\ref{tab:top-10-subreddits}, we report the top 10 subreddits with the highest number of submissions and comments, which seem rather ``science-oriented.''
This does not take into account engagement; are e-Prints posted as part of large, active discussions, or are they part of smaller conversations?
We define engagement as the number of comments the Reddit submission receives; thus, it is quantified by counting the number of comments in each submission in our dataset.
A submission with a large number of comments likely relates to high user engagement.
The top five Reddit submissions with the largest number of comments are reported in Table~\ref{tab:top-five-subreddits-submissions}.
({\em NB:} we replace the submission ID with (A), (B), etc.)

We find both science-oriented subreddits (e.g., r/science and r/COVID19) and generic ones (e.g., r/explainlikeimfive and r/AskReddit), as well as some controversial communities (e.g., r/The\_Donald, r/conspiracy).

\descr{4chan} We also count the number of posts in /pol/ and present the top five /pol/ threads with the largest number of links in Table~\ref{tab:top-five-pol-threads}.
We find /pol/ threads about COVID-19 (i.e., White House Chinese Virus Press Briefing \#2 and threads with `CVG' prefix).
Our findings show many COVID-19-related threads, but also threads about U.S. politics (i.e., 2020 Presidential Election Results, PTG: It's Time To Get Spooky Edition and PBG: Epic Victory Edition), ``The Storm'' conspiracy~\cite{NyMag2017} (i.e., CBTS: \#191 and CBTS: \#192 Meme Are Media), data science (i.e., Hi Data Scientist Here), and current events (i.e., Happening).

\begin{table}[t!]
  \centering
  \setlength{\tabcolsep}{4pt}
  \small
  \begin{tabular}{lr|lr}
      \toprule
      \multicolumn{2}{c|}{\textbf{\em Submissions}} & \multicolumn{2}{c}{\textbf{\em Comments}} \\
      \midrule
      \textbf{Subreddit}      & \textbf{$\mathbf{N}$}  & \textbf{Subreddit} & \textbf{$\mathbf{N}$} \\
      \midrule
      r/MachineLearning       & 3,762            & r/MachineLearning       & 7,643 \\
      r/statML                & 3,552            & r/askscience            & 3,531 \\
      r/ds\_links             & 1,571            & r/Physics               & 3,091 \\
      r/silky                 & 1,151            & r/math                  & 2,467 \\
      r/reinforcementlearning & 1,040            & r/science               & 2,137 \\
      r/cs\_theory            & 928              & r/space                 & 1,089 \\
      r/TopOfArxivSanity      & 877              & r/Physics\_AWT          & 823 \\
      r/science               & 852              & r/AskPhysics            & 637 \\
      r/Physics               & 811              & r/reinforcementlearning & 563 \\
      r/PlanetExoplanet       & 786              & r/ScienceUncensored     & 518 \\
      \bottomrule
  \end{tabular}
  \caption{Top 10 subreddits with e-Print links.}
  \label{tab:top-10-subreddits}
\end{table}

\begin{table*}[t!]
  \centering
  \small
  \begin{tabular}{lll}
    \toprule
    \textbf{arXiv} & \textbf{bioRxiv} & \textbf{medRxiv} \\
    \midrule
    r/technology(A) (2,915)         & r/IAmA(A) (10,988)            & r/pics(A) (3,168)     \\
    r/The\_Donald(A) (2,369)        & r/AskReddit(A) (8,545)        & r/conspiracy(A) (999)  \\
    r/The\_Donald(B) (1,727)        & r/slatestarcodex(A) (1,053)   & r/newjersey(A) (361)   \\
    r/explainlikeimfive(A) (1,291)  & r/science(B) (646)            & r/COVID19(A) (288)      \\
    r/science(A) (1,267)             & r/science(C) (459)            & r/conspiracy(A) (255)  \\
    \midrule
    \textbf{ChemRxiv} & \textbf{EarthArxiv} & \textbf{viXra} \\
    \midrule
    r/coronabr(A) (2,322)     & r/worldnews(A) (8,966)            & r/science(G) (299) \\
    r/Coronavirus(A) (1,732)  & r/worldnews(B) (6,941)            & r/conspiracy(B) (169)  \\
    r/science(D) (697)       & r/neoliberal(A) (5,212)           & r/viXra\_revA(A) (154)  \\
    r/COVID19(B) (558)       & r/science(E) (2,162)              & r/badmathematics(A) (105)         \\
    r/medicine(A) (506)      & r/science(F) (365)               & r/Physics(A) (105) \\
    \bottomrule
  \end{tabular}
    \caption{Top five Reddit submissions with at least one e-Print link. (NB: (A), (B), etc. replace the real submission ID.)}
  \label{tab:top-five-subreddits-submissions}
\end{table*}

\begin{table*}[t!]
  \centering
  \small
    \setlength{\tabcolsep}{2pt}
  \begin{tabular}{lll}
    \toprule
    \textbf{arXiv} & \textbf{bioRxiv} & \textbf{medRxiv} \\
    \midrule
    2020 Presidential Election Results (1,001) & CVG: \#257 Tears In The Rain (628) & CVG: Official Coronavirus General (1,215) \\
    CBTS: \#191 (483)  & CVG: \#295 Burn It Down Edition (583) & CVG: \#295 Burn It Down Edition (583)  \\ 
    CBTS: \#192 Meme Are Media (472)  & CVG: \#324 Only God Can Save Us (569) & CVG: \#324 Only God Can Save Us (569)  \\
    What's Wrong With Civic Nationalism (471)  & CVG: \#132 Batshit Edition (558) & CVG: \#325 The Commie Plague (553)  \\
    CVG: \#639 Burger Edition (455)  & CVG: \#325 The Commie Plague (553) & CVG: \#461 Let The Bodies Hit (552)  \\
    \midrule
    \midrule
    \textbf{ChemRxiv} & \textbf{PsyArXiv} & \textbf{viXra} \\
    \midrule
    White House Chinese Virus Press Briefing \#2 (393) & PTG: It's Time To Get Spooky Edition (354) & Happening (344) \\
    CVG: \#3084 (386) & Hi Data Scientist Here (323) & CVG: \#2825 Status Lockdown (342) \\
    CVG: \#3080 (386) & - & PBG: Epic Victory Edition (330) \\
    CVG: \#3000 (385) & - & CVG: \#5087 (328) \\
    CVG: \#3062 (382) & - & PBG: Purple (328) \\
    \bottomrule
  \end{tabular}
    \caption{Top five /pol/ threads with at least one e-Print link. \emph{Note: CBTS stands for `Calm Before The Storm', CVG stands for `Corona Virus General', PTG stands for `President Trump General', and PBG stands for `President Biden General.'}}
  \label{tab:top-five-pol-threads}
  \vspace{-0.1cm}
\end{table*}

\descr{{Main Takeaways}}
We find non-negligible evidence of e-Prints being linked on Reddit and /pol/.
The presence is found in science enthusiasts communities (e.g., r/science) and general audience communities with various interests (e.g., The\_Donald and /pol/).
On /pol/, the top five most engaged threads with bioRxiv and medRxiv links are about COVID-19.
Coincidently, there is a surge of /pol/ posts with these links in the early months of the COVID-19 outbreak, between late January and mid of March 2020~\cite{WHOTimeline}.
This also mirrors the explosion of e-Prints in the early months of the COVID-19 outbreak as per previous studies~\cite{Majumder_2020,Vlasschaert_2020,Kwon_2020,Fleerackers_2021,Fraser_2021}.

\subsection{What kind of e-Prints are linked to Reddit and /pol/?}

\descr{Overview} Although we can get a coarse understanding of what kind of e-Prints are linked in the discussions based on the server they are published on, and we can also rely on the various categories defined by the server, we would ideally shed light on what individual papers are about.
Thus, to discover the kind of e-Prints linked to Reddit and /pol/, we use the Top2Vec algorithm to build a topic model from a set of e-Prints of each server in our dataset; specifically, arXiv, viXra, and bioRxiv on Reddit and medRxiv, bioRxiv, and arXiv on /pol/ (see Table~\ref{tab:dataset-statistic}).

\descr{Top2Vec} More precisely, the algorithm creates topic embeddings of the e-Prints and uses them to discover a high-level summary of information (i.e., the topic) in a set of e-Prints~\cite{Angelov_2020}.
Top2Vec uses Doc2Vec to build the document embedding of each e-Print and Word2Vec to vectorize the words into word embeddings which are then used to measure the distance between the document and word vectors.
The dense area between them indicates a common topic to the e-Prints.

\descr{Topics} We keep the maximum number of topics to 10 topics.
This allows us to qualitatively review the representative documents in each topic and manually assign topic labels on the generated topics.
The output of the Top2Vec algorithm with the assigned topic labels is reported in Tables~\ref{tab:reddit-e-Print-topics} and~\ref{tab:pol-e-Print-topics}, respectively, for Reddit and /pol/.

\begin{table}[t!]
  \centering
  \small
    \setlength{\tabcolsep}{1.5pt}
  \begin{tabular}{lrl}
    \toprule
      \textbf{Server} & \multicolumn{2}{l}{\textbf{Topics}} \\
    \midrule
      arXiv & 1. & Computational Theory and Algorithms\\
& 2. & Deep Learning on NLP\\
& 3. & Classification, Clustering and Ranking Algorithms\\
& 4. & Computational Logic and Theorem\\
& 5. & Neural Network and Neuro-science\\
& 6. & Optoelectronic and Quantum Theory\\
& 7. & Video and Image Computational Analysis\\
& 8. & Space and Planetary Study\\
& 9. & Supernova in Astrophysics\\
& 10. & Cloud and High-Performance Computing \\
      bioRxiv & 1. & Genome Sequencing Study\\
& 2. & Cognitive and Neuro-imaging Study\\
& 3. & Genetic Study\\
&  4. & Ancestry Study \\
      viXra & 1. & Mathematical Theorem Proving\\
&  2. & Stellar Object Study \\
    \bottomrule
  \end{tabular}
    \caption{Topics of e-Print links on Reddit.}
  \label{tab:reddit-e-Print-topics}
\end{table}

\begin{table}[t!]
  \centering
  \small
      \setlength{\tabcolsep}{3pt}
  \begin{tabular}{lrl}
    \toprule
      \textbf{Server} & \multicolumn{2}{l}{\textbf{Topics}} \\
    \midrule
      arXiv & 1. & A.I. and Deep Learning\\
& 2. & Astronomy - Space and Universe\\ 
& 3. & Solar and Planetary Objects\\
& 4. & Climate Change\\
& 5. &  Atomic Particles \\
bioRxiv & 1. & COVID-19\\
& 2. & Population Genomic \\
& 3. & Genomic Ancestry \\
medRxiv & 1. & COVID-19 Symptoms Study\\
& 2. & Clinical Study on COVID-19\\
& 3. & Early Study on COVID-19 Outbreak\\
& 4. & Immunity and Antibody Study to COVID-19\\
& 5. & Hydrochloroquine and Invermectiv Study\\
& 6. & Infection and Transmission Study of COVID-19\\
& 7. & COVID-19 Variants Study\\
& 8. & COVID-19 Infection Fatality Risk \\
    \bottomrule
  \end{tabular}
    \caption{Topics of e-Print links on /pol/.}
  \label{tab:pol-e-Print-topics}

\end{table}

On Reddit, we find topics in Computer Science (e.g., Computational Theory and Algorithms, and Deep Learning on NLP) and Physics (e.g., Optoelectronic and Quantum Theory, Supernova in Astrophysics) from arXiv.
Topics discovered from bioRxiv are about Genome, Neuroscience, and Genetics, while viXra includes topics about Maths and Physics, particularly Astro-Physics.
On /pol/, we also observe Computer Science topics, particularly about deep learning, and Physics (e.g., Solar and Planetary Objects, Climate Change, and Atomic Particles).
Two topic themes are discovered for bioRxiv: COVID-19 and Genome, while medRxiv e-Prints are only about COVID-19.

\descr{{Main Takeaways}} On both Reddit and /pol/, we find e-Prints from Computer Science and Physics.
On /pol/, COVID-19 dominates the e-Print links from bioRxiv and medRxiv, which is in line with the top five /pol/ threads being about COVID-19 (see Table~\ref{tab:top-five-pol-threads}) and the surge shown in Figure~\ref{fig:pol-posts-count}.

\subsection{How are e-Prints Discussed on Reddit and /pol/?}

\descr{Overview} To understand how e-Prints are positioned within the discussions, we examine the similarity between the discussion thread and the e-Print linked in the discussion, using document embeddings. 
Then, we qualitatively review some of the threads.
More precisely, we select the top threads from Reddit submissions and /pol/, and review the discussions.
We choose one thread from /pol/ and two threads from Reddit consisting of one thread with a general audience and one science-enthusiasts community thread.
We intentionally select those threads to show examples of how e-Prints are positioned in 1)~a thread of fringe community like /pol/, 2)~a thread of general audience community, and 3)~a thread with a specific interest like science.

\descr{Doc2Vec} 
As mentioned, we build document embeddings (i.e., document vectors) using Doc2Vec. 
This unsupervised algorithm learns pieces of text like sentences, paragraphs, or documents with varying lengths~\cite{Le2014}.
Then, we use cosine similarity to measure the similarity between the generated document embeddings.

\begin{figure*}[t!]
  \centering
  \begin{subfigure}{0.33\linewidth}
    \includegraphics[width=\linewidth]{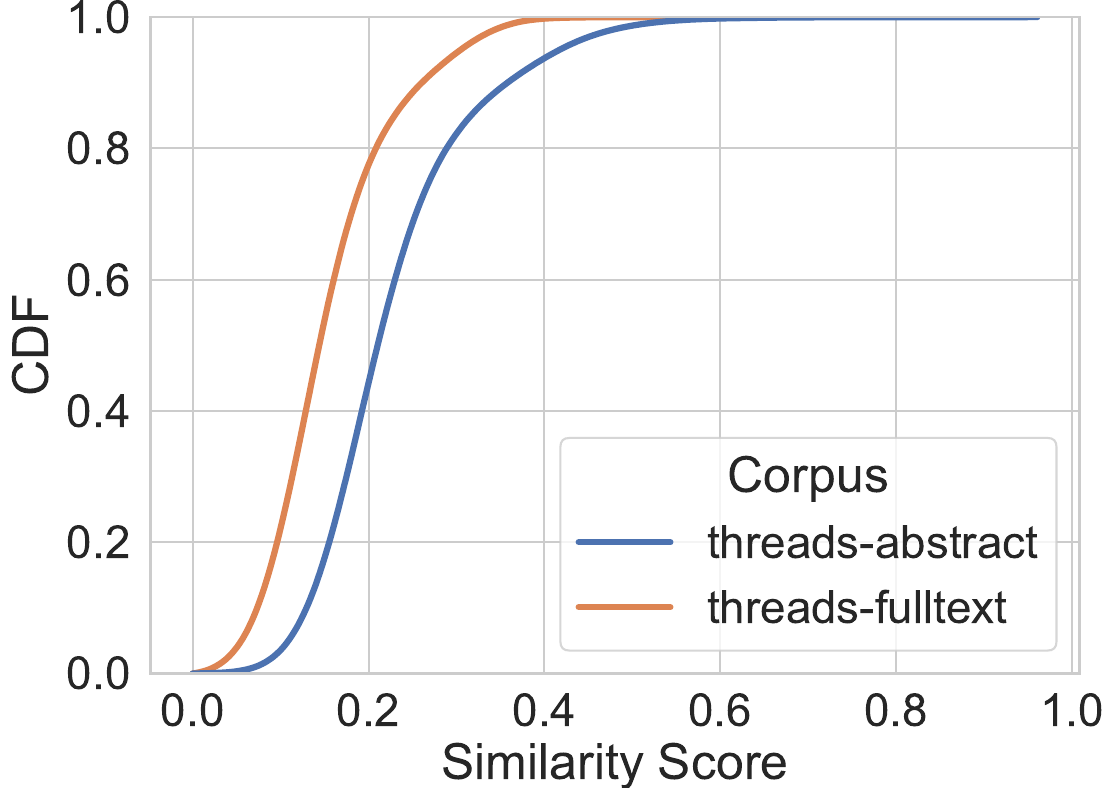}
    \caption{/pol/ - arXiv}
    \label{fig:pol-arxiv}
  \end{subfigure}
  \hfill
  \begin{subfigure}{0.33\linewidth}
    \includegraphics[width=\linewidth]{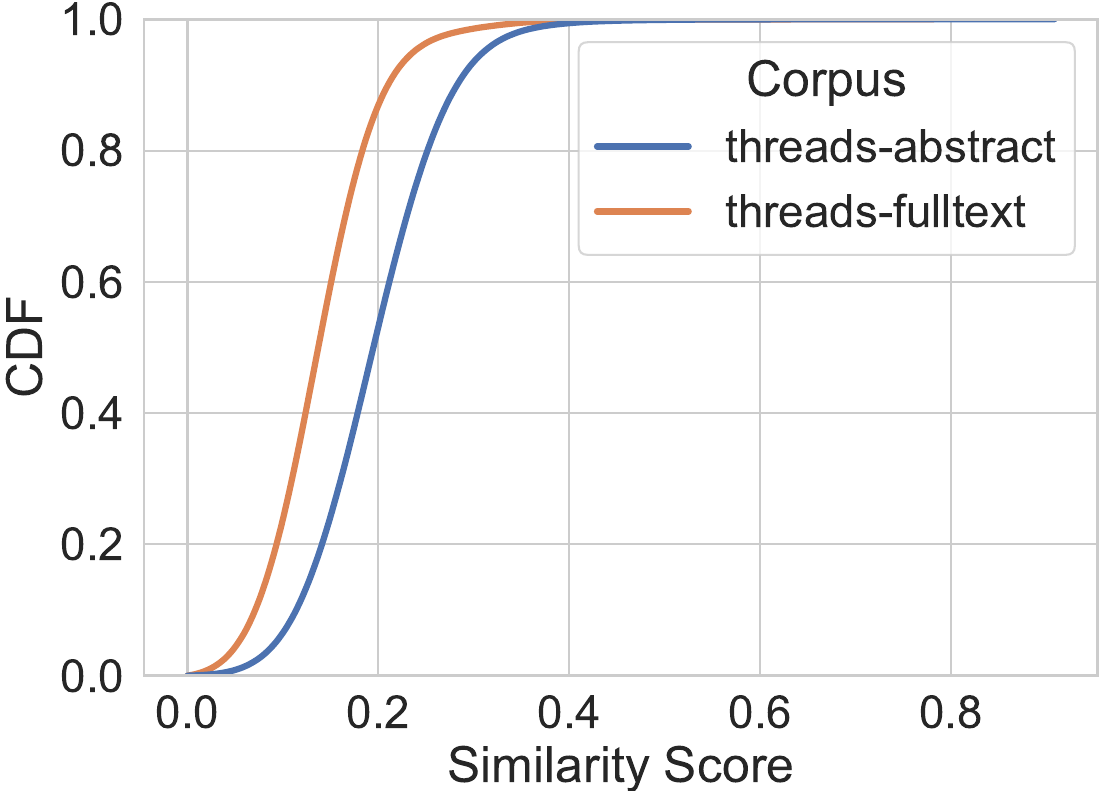}
    \caption{/pol/ - bioRxiv}
    \label{fig:pol-bioRxiv}
  \end{subfigure}
  \hfill
  \begin{subfigure}{0.33\linewidth}
    \includegraphics[width=\linewidth]{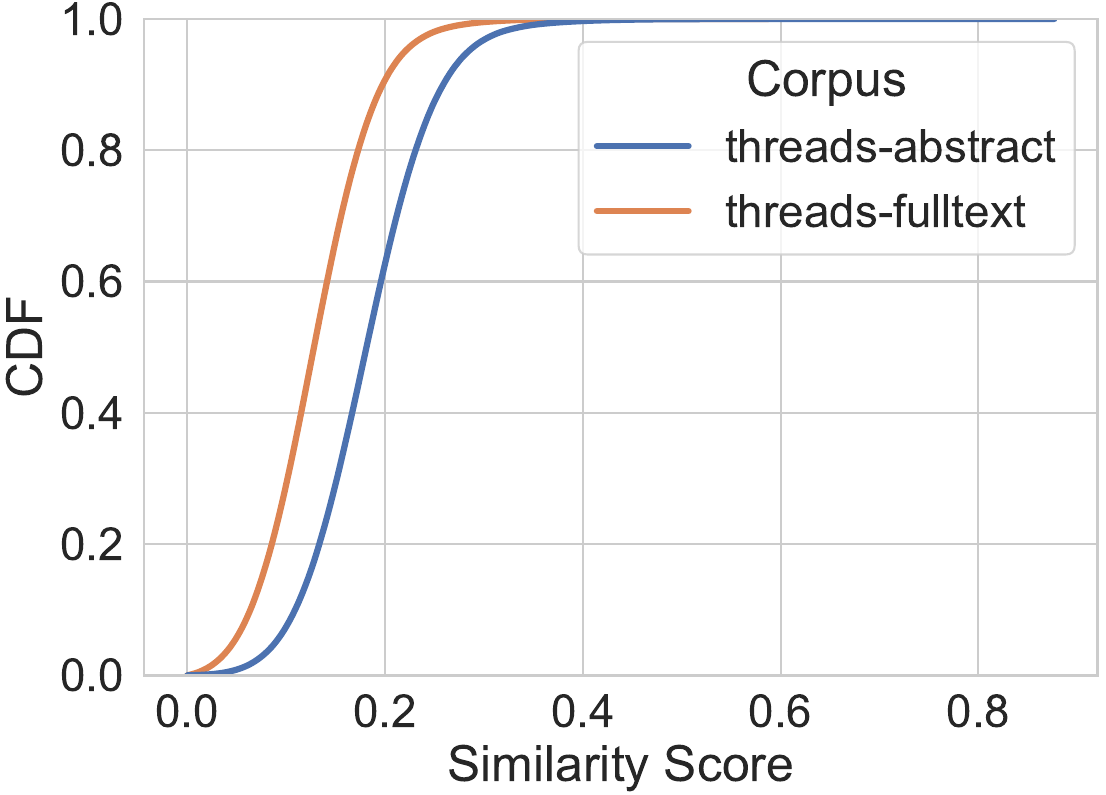}
    \caption{/pol/ - medRxiv}
    \label{fig:pol-medRxiv}
  \end{subfigure}
  \medskip
  \begin{subfigure}{0.33\linewidth}
    \includegraphics[width=\linewidth]{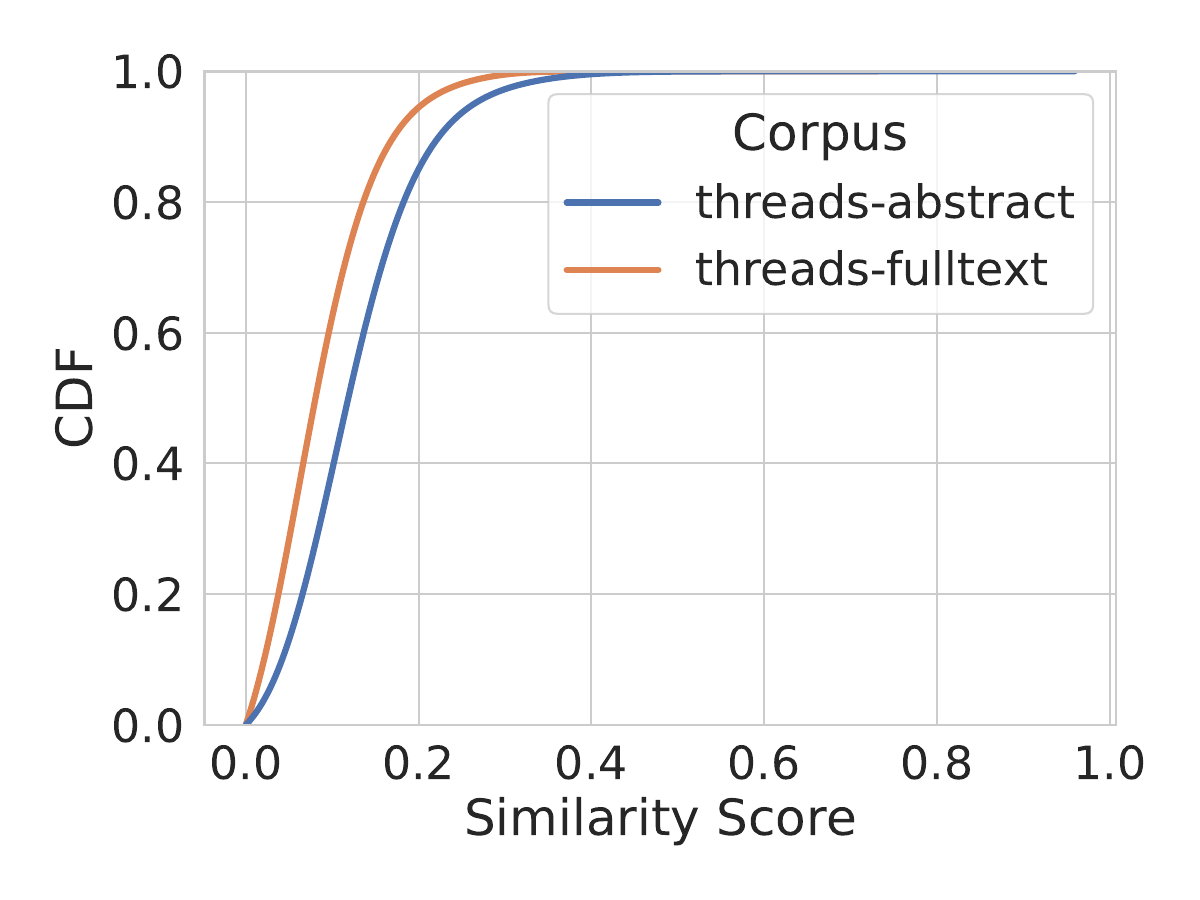}
    \caption{reddit - arXiv}
    \label{fig:reddit-arxiv}
  \end{subfigure}
  \hfill
  \begin{subfigure}{0.33\linewidth}
    \includegraphics[width=\linewidth]{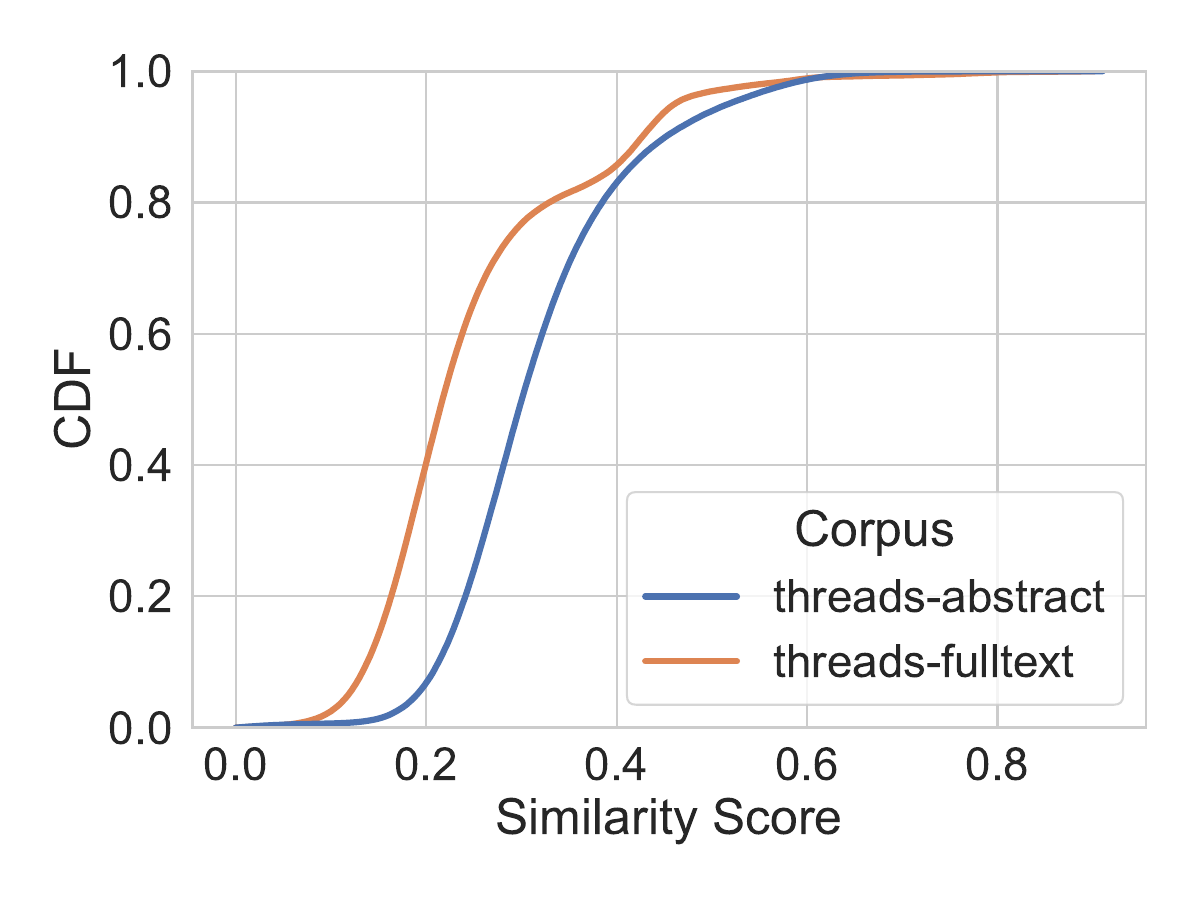}
    \caption{reddit - bioRxiv}
    \label{fig:reddit-bioRxiv}
  \end{subfigure}
  \hfill
  \begin{subfigure}{0.33\linewidth}
    \includegraphics[width=\linewidth]{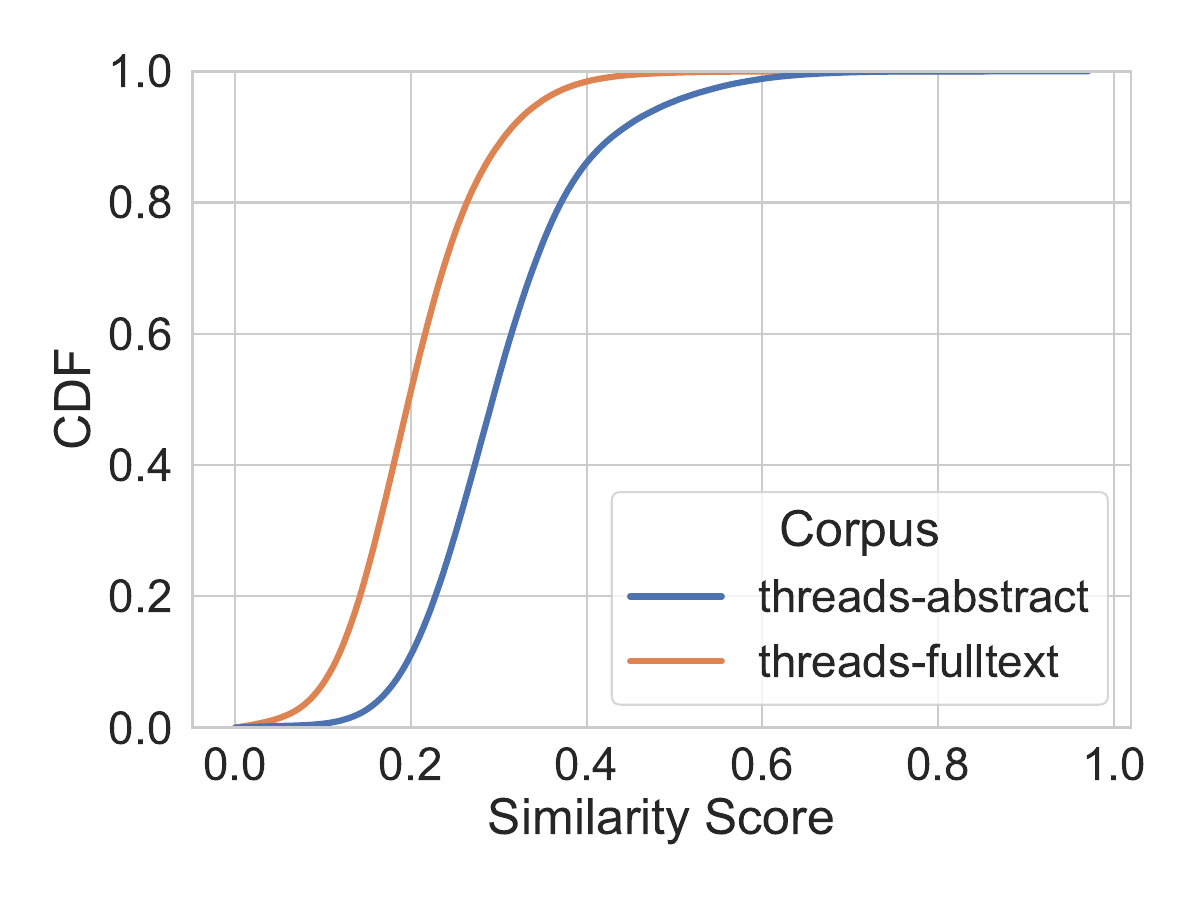}
    \caption{reddit - vixra}
    \label{fig:pol-vixra}
  \end{subfigure}
  \caption{The CDF of similarity score between each e-Print and the thread it appears in.}
  \label{fig:cdf-thread-eprint}
\end{figure*}

\descr{Similarity} In Figure~\ref{fig:cdf-thread-eprint}, we plot the similarity distributions between thread and e-Print document embeddings both 1)~between threads and e-Print abstracts and 2)~between threads and e-Print full texts.
The former tends to be higher.
The overall average similarity scores between the e-Print and the thread it appears in are rather low (<0.32); see Table~\ref{tab:similariy-eprint-thread}.

\begin{table}[t!]
  \centering
  \small
  \centering
    \begin{tabular}{lrr}
      \toprule
      \textbf{Server} & \textbf{Th-Abs (Mean)} & \textbf{Th-Full (Mean)} \\
      \midrule
      arXiv           & 0.122         & 0.076         \\  
      bioRxiv         & 0.315         & 0.247         \\
      viXra           & 0.302         & 0.202         \\
         \midrule
                 &\textbf{ (a) Reddit}         &         \\
               &          &          \\
      \toprule
      \textbf{Server} & \textbf{Th-Abs (Mean)} & \textbf{Th-Full (Mean)} \\
      \midrule
      arXiv           & 0.227         & 0.154         \\  
      bioRxiv         & 0.197         & 0.139         \\
      medRxiv         & 0.183         & 0.129          \\
      \midrule
                 & \textbf{(b)  /pol/}    &         \\
    \end{tabular}
  \caption{The average similarity score between each e-Print and the thread it appears in. \emph{Note: Th-Abs = Thread and Abstract; Th-Full = Thread and Full Text.}}
  \label{tab:similariy-eprint-thread}

\end{table}

\descr{Manual Review} Next, we manually examine threads to understand the positioning of e-Prints in discussions.
We select two threads each from Table~\ref{tab:top-five-subreddits-submissions} and Table~\ref{tab:top-five-pol-threads}.
We purposely select general audience threads and a more-narrow audience like science-enthusiasts threads:
\begin{compactenum}
\item r/The\_Donald(A), a subreddit representing the general audience thread
\item CVG: \#132 Batshit Edition, a thread from /pol/ representing the general audience thread
\item r/science(G), a science-enthusiast subreddit
\item Hi Data Scientist Here, a /pol/ thread for data scientists.
\end{compactenum}

\descr{{r/The\_Donald(A) - Quarantine Update Transparency Report}}
The r/The\_Donald subreddit was known for its support to U.S. President Trump and a well-documented set of content policy violations that led to its quarantining, eventual ban, and migration to a new platform~\cite{Ribeiro_2020}.
This submission discusses the content moderation decisions about r/The\_Donald itself, specifically, a denied appeal to lift the quarantine status that Reddit moderators had put in place.
In the discussion, r/The\_Donald moderators reported the quarantine update, and the moderators responded to the denied appeal by Reddit moderators.
In part of the most recent appeal, r/The\_Donald moderators noted their effort to reduce policy-violating content, e.g., racist content.
One of their arguments is referencing an e-Print~\cite{Zannettou_2018} claiming the rest of Reddit is 50\% more racist than r/The\_Donald, as paraphrased in the following:
\begin{quote}
  \emph{According to a 2018 meme study~\cite{Zannettou_2018}, it craftily said r/The\_Donald produced more racist memes than other subreddits.
    However, when they measured the so-called ``racist'' memes proportional to the total memes on r/The\_Donald (0.4\%), the rest of Reddit is 50\% more racist (0.6\%).}
\end{quote}

However, our examination of the linked e-Print~\cite{Zannettou_2018} shows racist memes are pervasive in fringe communities like r/The\_Donald and /pol/, and r/The\_Donald was the most \emph{efficient} community for propagating memes to other communities.
Therefore, it contradicts with the argument made by the r/The\_Donald moderators.
This example shows a misinterpretation of research findings in the linked e-Print to appear to support the statement made by The\_Donald moderators.

\descr{{/pol/ Corona Virus General: \#132 Batshit Edition}}
This thread is one of several in /pol/ that discusses the COVID-19 pandemic (see Table~\ref{tab:top-five-pol-threads}).
The original poster started the thread with a statement \emph{``Full Lockdown and Quarantine on 21 Major Cities/Provinces''} followed by a list of numbers of 1)~confirmed infected countries, 2)~unconfirmed/suspicious infected countries, and 3)~confirmed cases in the U.S.
This thread was created on January 26, 2020, when the WHO assessed whether the outbreak in Wuhan, China constituted a global public health emergency or not~\cite{WHOTimeline}.
Thus, the COVID-19 discussion was still in its infancy.
Such infancy is shown in posts like \emph{``I'm wondering how it's comparable to H1N1''} and \emph{``so like 80 people died in China. Why is this any cause of concern?''}.

In the following conversation, it discusses the predicted death count, and one user puts a link to a bioRxiv e-Print~\cite{Zhao2020} presenting the first study to quantify the basic reproduction number ($R_0$) of COVID-19 in the early phase of the outbreak.
The data used in the study comprises all laboratory-confirmed cases released by the Wuhan Municipal Health Commission and National Health Commission of China from January 10 to January 22, 2020.
The bioRxiv e-Print was uploaded to the server on January 24, 2020, and the post was submitted on January 26, 2020. 
The post is paraphrased as follows:
\begin{quote}
    \emph{The latest $R_0$ is 5.47 according to~\cite{Zhao2020} which means CANCEL ALL FLIGHTS NOW level pandemic bad}
\end{quote}

\noindent One response gives a counterargument about the exponential growth of the reproduction number mentioned in the e-Print:
\begin{quote}
	\emph{The Chinese CDC released a report saying the reproduction number is between 2.5 - 2.9}
\end{quote}

\noindent Another response is worried by the exponential growth:
\begin{quote}
  	\emph{The reproduction growth from 3.30 to 5.47 means that COVID-19 is an incredibly aggresive virus with 14 days of incubation. This is a REAL serious global pandemic. Gather your weapons, food, water and masks. Let's hope that the vaccine will be ready within 3 months.}
\end{quote}

\noindent This example shows that early research findings influence layperson judgement, particularly in speculating COVID-19 infection rate.

\descr{{r/science(G) - Particle Physicists Confirm Arrow of Time}}
The r/science subreddit is a community to share and discuss new scientific research.
The original poster of this thread shared a link to a news blog from Nature~\cite{Reich2012} about the direct measurement result to confirm that time does not run the same forwards as backwards.
Our review finds deep discussion about the result between science enthusiasts who lack understanding about the subject and those who do.

Before the post that referenced to the e-Print, one Redditor posted their dissatisfaction with not being able to download the original paper mentioned in the Nature news blog.
In reply to that, a Redditor posts a link to a viXra e-Print~\cite{Tooker2012}, which supposedly the original idea of the experiment ran in~\cite{Reich2012}.
An immediate response to the post says the Redditor who posted the viXra e-Prints is a spammer, as shown in the following paraphrased quote:
\begin{quote}
    \emph{You spammed my department with the non-sense paper. You even referenced the Mayan calendar in your research. You are delusional to think that you can push this pseudoscientific non-sense to everybody so that it will make you a successful scientist.}
\end{quote}

\noindent The discussion continues questioning viXra e-Print's relation to the original paper and the validity of the e-Print.
One Redditor explains the reason that the viXra e-Print raises a red flag:
\begin{quote}
    \emph{People see your paper as a red flag because (1) the abstract shows something vague and alternative and nothing concrete, and (2) there is no introduction in your paper and immediately jumping to long derivation of unfamiliar notations.  Therefore, this gives me a clue to not read your paper.}
\end{quote}

This example shows a debate made between a group of science enthusiasts with a particular person who shared an e-Print from viXra.
In this case, it is a pseudoscience e-Print which causes it to be questioned and rejected by science enthusiasts in a science-oriented community like r/science.

\descr{{/pol/ Hi Data Scientist Here}}
The thread name implies that the thread discussed data science-related topics between whom we assume to be data scientists.
The original poster states that studies are lying in using data in which most \emph{normies}, a derogatory slang referring to normal people, are ignorant of statistics and data manipulation.
Moreover, they believe that most studies are \emph{psyops}, a term referring to psychological operations that influence the opponent's state of mind.
They take an example of election prediction with a hidden agenda, where collected data is analyzed to support it.

In reply to the original poster, users are aware of the data manipulation practice, mainly using statistical modeling to support the pursued agenda.
One user argues that validating data in many studies is difficult as most people do not have the tools or the know-how to analyze the data, and occasionally the dataset itself is unavailable to the public.

We find one user refers to an e-Print stored in PsyArxiv with a title of \emph{``Many Labs 2: Investigating Variation in Replicability Across Sample and Setting''}~\cite{Klein2018} and says \emph{``most classic psychological experiments failed in replicability test.''}.
However, the user does not elaborate on the paper's findings.
Therefore, we assumed the user believed they would fail to replicate the existing studies even though they have the data to validate.
Our review of the e-Print shows that it examines variation in effect magnitudes across the sample and experiment setting of 28 classic and contemporary published findings.
The study aims to probe the variability of psychological effects across samples by testing them across a broad range of contexts. 
This allows an understanding of the extent to which a given psychological finding represents general features of the human mind.

This example shows a PsyArxiv e-Print being used as a reference to generalize that most (psychological) experiments are irreproducible.
We do not dismiss that some experiments and studies are irreproducible.
Still, the user quickly concluded the paper without explaining the objective of the study and the argument of why they are irreproducible.

\descr{{Main Takeaways}}
Overall, we observed low similarity between e-Print and the threads it appears in.
Upon manual examination, we uncovered a number of interesting insights.
For instance, we found a thread that misinterpreted the research findings in an e-Print to support the argument made in the discussion.
On another thread, a user quickly generalized that most experiments are irreproducible by citing an e-Print without explaining the study's findings and context.
Also, early research findings about COVID-19 reproduction numbers based on laboratory-confirmed cases were often used to speculate the COVID-19 infection rate.
Finally, we shed light on a group of science enthusiasts identifying pseudoscience e-Print and questioning the validity of the research findings.

\section{Discussion \& Conclusion}
This paper examined the presence of e-Prints on Reddit and /pol/.
Using over 14 years of Reddit data and over four years of /pol/ data, we looked for posts containing links to e-Prints from eight servers -- i.e., arXiv, bioRxiv, medRxiv, ChemRxiv, viXra, PsyArXiv, SocArxiv, and EarthArxiv.

\descr{Summary of quantitative results} On Reddit, arXiv papers are shared the most, with 50,904 e-Prints in total, followed by viXra (975), bioRxiv (781), ChemRxiv (69), medRxiv (60), and EarthArxiv (24).
On /pol/, medRxiv hosted the most shared e-Prints with 719 in total, followed by bioRxiv (676), arXiv (427), viXra (27), PsyArXiv (17) and ChemRxiv (5)
We overall found e-Prints present in science enthusiast communities (e.g., r/science) and general audience communities of various interests (e.g., r/The\_Donald and /pol/).

Our topic analysis found that Computer Science and Physics (arXiv) e-Prints are the most shared.
We also found many e-Prints about COVID-19 (from bioRxiv and medRxiv) on /pol/.
This is in line with the surge of /pol/ posts with links to bioRxiv and medRxiv between late January and mid-March 2020, corresponding to an overall increase in the number of COVID-19 related e-Prints~\cite{Majumder_2020,Vlasschaert_2020,Kwon_2020,Fleerackers_2021,Fraser_2021}.

Social media discussions that linked to e-Prints were not very similar to the contents of the e-Prints themselves.
Interestingly, the similarity between discussions and the \emph{abstracts} of e-Prints tended to be higher than between discussions and the full contents of the e-Prints.
This likely indicates that either e-Prints are only a small part of the overall discussion where they are referenced or that users simply do not read past the abstract.

\descr{In-depth analysis} We then performed a more nuanced, qualitative analysis.
We observed research findings of e-Prints being misinterpreted to support a particular argument in a discussion.
We also found a case where a user cited an e-Print without adequately explaining the findings and context of the study to argue that most experiments are irreproducible.
In another example, early findings about COVID-19 reproduction numbers were used to speculate about COVID-19's infection rate.
Finally, we saw a group of science enthusiasts identifying and debunking a pseudoscience e-Print shared by a different user.

\descr{{Limitations}} Our work only focuses on the presence of e-Prints from eight different servers on Reddit and /pol/.
However, while these servers are open-access repository servers for scientific articles that do not require peer review, there is at least a basic system to screen out completely random content.
There are also open-access repositories of peer-reviewed scientific articles, like PLoS, Frontiers, and MDPI, which we do not study
Further, we made no distinction between e-Prints that have been peer-reviewed and published and e-Prints that have not made it past peer review.
Finally, we focused solely on understanding the use of e-Prints on Reddit and /pol/; future work could explore other Web communities as well as other scientific metrics (e.g., scientometric and bibliometrics).

\descr{{Implications}}
Our findings show that e-Prints are being heavily disseminated on Reddit and /pol/ where users with different levels of expertise may interpret their content differently.
Consequently, incorrect interpretations or low-quality papers may become ``gold standards'' to some.
While we acknowledge the e-Print model's advantages (e.g., rapid and open distribution), we are also mindful of its drawbacks and the potentially dangerous implications for the scientific community at large, including but not limited to those highlighted by our work.

Our study shows research findings \emph{are} used by laypeople, and we as scientists have no control over \emph{how} they will be used.
This is exacerbated by the open access, yet potentially not peer-reviewed nature, of e-Prints.
In short, the traditional model of scientific publishing is a form of gatekeeping, but this gatekeeping \emph{does} serve a purpose.
At the very least, peer review provides some guarantee that qualified scientists have read and at least not entirely rejected the work in the paper.
However, it also comes with a price: readers must wait for a paper to make it through peer review and then \emph{pay} for access.
The peer review system helps ensure that the scientific community is self-policing and mitigates potential misinterpretation by non-scientists since relatively few people (i.e., scientists) are interested in paying for an article.

Even though we only study e-Prints servers that do not impose a peer review system, we believe this issue applies to open-access repositories of peer-reviewed scientific articles as well.
In the case of open-access repositories of peer-reviewed scientific articles, the responsibility to have the articles be openly accessible lies to the authors instead of the readers.
This is just a shift of the burden of who pays the publisher.

Our results also make it clear that the audience for scientific work is far greater than the scientific community itself.
Therefore, this implies we should consider additional ways to communicate research findings to a broader audience.
One solution might be to explain our research in ``simple English'' in outlets like The Conversation, which provides a direct conduit for scientists to speak to laypeople about our work.
We encourage researchers to take this to heart and consider budgeting for not just publication fees and travel to conferences but for \emph{direct} dissemination to the public in their proposals to funding agencies.
This is a simple and straightforward first pass to help mitigate misinterpretation, with the added benefit of increasing positive public science engagement.

\descr{Acknowledgements}
Satrio Yudhoatmojo would like to thank DIKTI-funded Fulbright Grants for supporting his doctoral study at Binghamton University.
This research was supported by the National Science Foundation under Grant No.~IIS-2046590.
We also would like to thank Media Ecosystems Analysis Group and the Bill \& Melinda Gates Foundation for their generous support.
\small
\bibliographystyle{abbrv}

%
%
%

%

\end{document}